\documentclass{elsart}

\usepackage{graphicx}
\usepackage{amssymb}

\begin{document}
\begin{frontmatter}

\title{Competing orders and the doping and momentum dependent 
quasiparticle excitations in cuprate superconductors\thanksref{support}}
\thanks[support]{Work supported by the National Science Foundation through Grants \#DMR-0405088.}

\author{A.D. Beyer},
\author{C.-T. Chen},
\author{M.S. Grinolds},
\author{M.L. Teague},
\author{N.-C. Yeh\corauthref{cor}}
\corauth[cor]{Corresponding author.}
\ead{ncyeh@caltech.edu}

\address{Department of Physics, California Institute of
Technology, Pasadena, CA 91125}

\begin{abstract}
The low-energy quasiparticle excitations in hole- and electron-type cuprate 
superconductors are investigated via both experimental and theoretical means. 
It is found that the doping and momentum dependence of the empirical low-energy 
quasiparticle excitations is consistent with a scenario of coexisting competing 
orders and superconductivity in the ground state of the cuprates. This finding, 
based on zero-field quasiparticle spectra, is further corrobarated by the spatially 
resolved vortex-state scanning tunneling spectroscopy, which reveals pseudogap-like 
features consistent with a remaining competing order inside the vortex core upon the 
suppression of superconductivity. The competing orders compatible with 
empirical observations include the charge-density wave and spin-density wave.
In contrast, spectral characteristics derived from incorporating the 
$d$-density wave as a competing order appear unfavorable in comparison with 
experiments. 
\end{abstract}

\end{frontmatter}

\section{Introduction}
\label{Intro}
The physical origin of the pseudogap (PG) phenomena~\cite{LeePA06} and the 
apparent differences in the low-energy excitations of the hole- and electron-type 
cuprates~\cite{Renner98,Timusk99,Kleefisch01,ChenCT02,Yeh05b} are important and 
unresolved issues in high-temperature superconductivity. Empirically, the PG 
phenomena occur at two different energy scales~\cite{LeePA06}. One energy scale
is a high-energy PG existent in both electron- and hole-type 
cuprates~\cite{Onose01,HwangJ06,Matsui05,Yeh07}. The other energy scale 
is a low-energy PG coexisting with superconductivity (SC) at low temperatures, 
which survives above the superconducting transition $T_c$ and is manifested by 
the suppressed quasiparticle density of states (DOS)~\cite{Renner98,Timusk99}. 
This PG has only been observed in hole-type cuprates and is found to correlate 
with the onset of the Nernst effect~\cite{LeePA06,WangY06,WangY05}.  
Our recent studies of the quasiparticle spectral density function and the DOS 
of various optimally doped cuprates~\cite{ChenCT07} have revealed that 
the occurrence (absence) of the low-energy PG in hole- (electron-) type cuprate 
superconductors and the dichotomy in quasiparticle coherence~\cite{Matsui05,ZhouXJ04} 
can be quantitatively understood as the result of a ground state consisting of coexisting 
competing orders (COs) and SC~\cite{Zhang97,Kivelson03,Demler01,Polkovnikov02,Chakravarty01,ChenCT03,ChenHY05b} 
with the presence of finite quantum fluctuations~\cite{Yeh05b,Zapf05,Beyer07}. 
The objective of this work is to expand on our previous work~\cite{ChenCT07} to 
further investigate how the coexistence of COs and SC may influence the doping 
and momentum dependence of the low-energy quasiparticle excitations in different cuprates. 
Specifically, we compare empirical quasiparticle spectra obtained from scanning 
tunneling spectroscopy and ARPES (angle-resolved photoemission spectroscopy) 
data in zero field with calculated quasiparticle DOS and spectral density functions. 
We focus on three COs: the charge-density wave (CDW)~\cite{Kivelson03}, 
spin-density wave (SDW)~\cite{Schrieffer89,Demler01,Polkovnikov02}, and 
$d$-density wave (DDW)~\cite{LeePA06,Chakravarty01}. In addition, we perform
spatially resolved quasiparticle tunneling spectra on both hole- and electron-type
cuprates in the vortex state to investigate whether the spectral 
characteristics inside a vortex core are consistent with the presence 
of additional orders upon the suppression of SC. We find that the vortex-core 
states of both types of cuprates exhibit PG-like spectra, which differ from either 
the broad zero-bias bound states or flat conductance inside of the vortex 
core of conventional superconductors~\cite{Fischer07}. These finite-field results 
may be attributed to a remaining CO upon the suppression of SC, and are
therefore corroborative of the notion of coexisting COs and SC in the cuprates. 

\section{Theoretical modeling}
\label{Model}

To investigate the effect of coexisting COs and SC on the low-energy excitations, 
we incorporate both COs and SC in the mean-field Hamiltonian ${\cal H}_{MF}$ and 
further include the quantum phase fluctuations associated with the coexisting CO and 
SC phases in the proper self-energy~\cite{ChenCT07}. Specifically, we first obtain 
the bare Green's function $G_0 (\textbf{k},\omega)$ associated with a mean-field 
Hamiltonian ${\cal H}_{MF} = {\cal H}_{\rm SC} + {\cal H}_{\rm CO}$, where 
\begin{equation}
{\cal H}_{\rm SC} = \sum _{\textbf{k},\sigma} \xi _{\textbf{k}} c^{\dagger} _{\textbf{k},\sigma} c_{\textbf{k},\sigma} 
- \sum _{\textbf{k}} \Delta _{\rm SC} (\textbf{k}) (c^{\dagger} _{\textbf{k},\uparrow} c^{\dagger}_{-\textbf{k},\downarrow}
+ c_{-\textbf{k},\downarrow} c_{\textbf{k},\uparrow})
\end{equation}
is the superconducting Hamiltonian for a given pairing potential 
$\Delta _{\rm SC} (\textbf{k})$, $\xi _{\textbf{k}}$
is the normal-state eigenenergy, and $\sigma = \uparrow , \downarrow$ refers to
the spin states. The mean-field CO Hamiltonian ${\cal H}_{\rm CO}$ with an energy 
scale $V_{\rm CO}$ and a wave-vector $\textbf{Q} _1$ for CDW, $\textbf{Q} _2$ for
disorder-pinned SDW and $\textbf{Q} _3$ for DDW can be expressed as 
follows~\cite{ChenCT07}:
\begin{eqnarray}
{\cal H}_{\rm CDW} &= \sum _{\textbf{k},\sigma} V_{\rm CDW} \left( 
c^{\dagger} _{\textbf{k},\sigma} c_{\textbf{k}+\textbf{Q}_1,\sigma} 
+ c^{\dagger} _{\textbf{k}+\textbf{Q}_1,\sigma} c_{\textbf{k},\sigma} \right) \qquad
\qquad \nonumber \\
{\cal H} ^{\rm pinned} _{\rm SDW} &= g^2 \sum _{\textbf{k},\sigma} V_{\rm SDW} \left( 
c^{\dagger} _{\textbf{k},\sigma} c_{\textbf{k}+\textbf{Q} _2,\sigma} 
+ c^{\dagger} _{\textbf{k}+\textbf{Q} _2,\sigma} c_{\textbf{k},\sigma} \right) 
\qquad \quad \nonumber \\
{\cal H}_{\rm DDW} &= \sum _{\textbf{k},\sigma} V_{\rm DDW} 
\left( \cos k_x - \cos k_y \right) \left( i c^{\dagger} _{\textbf{k}+\textbf{Q} _3,\sigma} 
c_{\textbf{k},\sigma} + h.c. \right) ,
\label{eq:Hco}
\end{eqnarray}
where the coefficient $g$ in ${\cal H}_{\rm SDW} ^{\rm pinned}$ represents the coupling 
strength between disorder and SDW~\cite{Polkovnikov02}, $\textbf{Q} _1$ is along 
the CuO$_2$ bonding direction $(\pi , 0)$ or $(0, \pi)$, 
$\textbf{Q} _2 = \textbf{Q} _1 /2$, and $\textbf{Q} _3$ 
is along $(\pi ,\pi)$~\cite{Chakravarty01}. Here we have assumed that the density waves 
are static because dynamic density waves can be pinned by disorder, and in the latter
case the momentum {\bf k} remains a good quantum number as long as the mean free path 
is much longer than the superconducting coherence length, a condition 
generally satisfied in the cuprates at low temperatures. We have also 
neglected the direct coupling of antiferromagnetic SDW to SC~\cite{Schrieffer89} 
in the Hamiltonian because the corresponding phase space contribution of 
the first-order SDW coupling to the DOS is too small in the doped 
superconducting cuprates, similar to the situation of negligible DDW 
coupling to the DOS in the doped cuprates, to be elaborated in the next section. 
We further note that the spectroscopic characteristics associated with either CDW 
or disorder-pinned SDW as the CO are similar in the charge sector, although the 
wave-vector of CDW is twice of that of SDW~\cite{Polkovnikov02}. 

To incorporate quantum phase fluctuations, we introduce a proper 
self-energy $\Sigma ^{\ast}$ in the zero-temperature zero-field limit. 
In this limit the longitudinal phase fluctuations dominate so that 
we can approximate $\Sigma ^{\ast}$ by the longitudinal part of the
one-loop velocity-velocity correlation $\Sigma ^{\ast} _{\ell}$~\cite{Kwon99a,Kwon01}: 
\begin{equation}
\Sigma ^{\ast}  _{\ell} \left( \textbf{k}, \omega \right) = \sum _{\textbf{q}} 
\left[ m \textbf{v} _g (\textbf{k}) \cdot \hat{\textbf{q}} \right] ^2 
C_{\ell} (\textbf{q}) G \left( \textbf{k} - \textbf{q}, \omega \right),
\label{eq:self}
\end{equation}
where the group velocity $\textbf{v}_g$ is given by 
$\textbf{v}_g = (\partial \xi _{\textbf{k}} / \partial k)/\hbar$
for $|\textbf{k}| \sim k_F$, and $C_{\ell} (\textbf{q})$ is 
a coefficient that measures the degree of quantum fluctuations as detailed in 
Refs.~\cite{ChenCT07,Kwon99a,Kwon01}, $\textbf{q}$ is the momentum of 
quantum phase fluctuations, and $\xi _{\textbf{k}}$ is given by realistic
bandstructures~\cite{ChenCT07}. Given $G_0 (\textbf{k}, \omega)$ and 
$\Sigma ^{\ast} _{\ell} (\textbf{k}, \omega)$ in $(4 \times 4)$ matrices 
with the basis $(c^{\dagger} _{\textbf{k} \uparrow} \ c _{-\textbf{k} \downarrow} 
\ c^{\dagger} _{\textbf{k}+\textbf{Q} \uparrow} \ c _{-(\textbf{k}+\textbf{Q}) \downarrow})$, 
we can derive the full Green's function 
$G (\textbf{k},\tilde \omega)$ through the Dyson's equation~\cite{ChenCT07}: 
\begin{equation}
G^{-1} (\textbf{k}, \tilde \omega) = G_0 ^{-1} (\textbf{k}, \omega) 
- \Sigma ^{\ast}  _{\ell} (\textbf{k}, \tilde \omega), 
\label{eq:Dyson}
\end{equation}
where $\tilde \omega = \tilde \omega (\textbf{k}, \omega)$ denotes the energy renormalized 
by the phase fluctuations~\cite{ChenCT07,Kwon99a}. The Dyson's equation in Eq.~(\ref{eq:Dyson}) 
can be solved self-consistently~\cite{ChenCT07} by first using the mean-field values of 
$\xi _{\textbf{k}}$ and $\Delta _{\rm SC}$ and choosing an energy $\omega$, then going 
over the momentum $\textbf{k}$-values in the Brillouin zone by summing over a finite 
phase space in $\textbf{q}$ near each $\textbf{k}$, and finishing by finding the corresponding 
fluctuation renormalized quantities $\tilde \xi _{\textbf{k}}$, $\tilde \omega$ and 
$\tilde \Delta _{\rm SC}$ until the solution to the full Green's function 
$G(\textbf{k}, \tilde \omega)$ converges with an iteration method~\cite{ChenCT07}. 
The converged Green's function yields the spectral density function 
$A( \textbf{k}, \omega ) \equiv - \rm{Im} \left[ G(\textbf{k}, 
\tilde \omega (\textbf{k},\omega)) \right] / \pi$ and the DOS ${\cal N} (\omega) 
\equiv \sum _{\textbf{k}} A( \textbf{k}, \omega )$.

\section{Comparison with empirical data of quasiparticle excitations}
\label{Exp}

\begin{figure}
\begin{center}
\includegraphics*[keepaspectratio=1,height=3in]{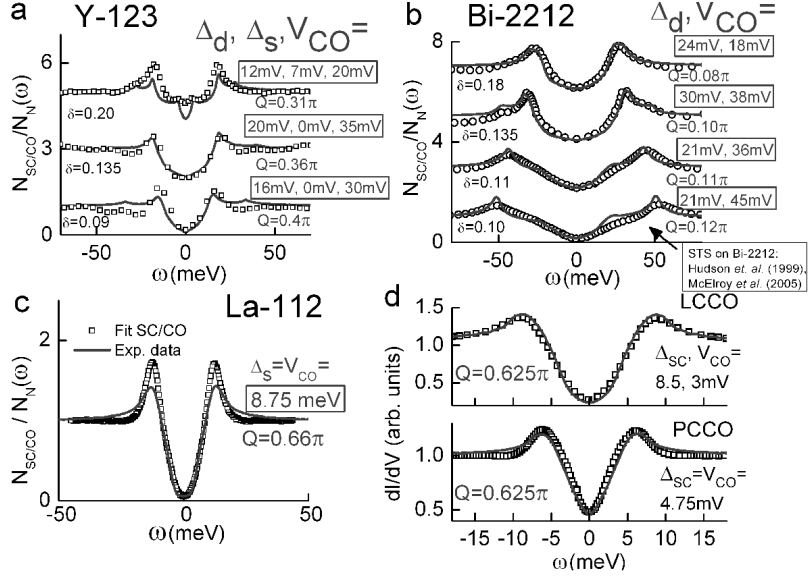}
\end{center}
\caption{(a) Comparison of normalized c-axis 
tunneling spectra on Y-123 of varying doping levels~\cite{Yeh01,Wei98} 
(symbols) with theoretical calculations (solid lines), where the
conductance of the spectra for different doping levels are slightly offset 
for clarity. The fitting parameters ($V_{\rm CO}$, $\Delta _{\rm SC}$) 
are found to be nearly independent of considering either bonding or anti-bonding
bands, and the $\textbf{Q}$-values listed are obtained by matching the Fermi level
of the antibonding band. Alternative fitting to the bonding band yields 
different Q-values of 0.66 $\pi$, 0.64 $\pi$ and 0.61 $\pi$ for $\delta$ = 0.09, 
0.135 and 0.20, respectively, and the fitted $\eta$ value is $\sim 5 \times 10^{-7}$
for $\delta$ = 0.09 and 0.135, and $\eta \sim 0$ for $\delta$ = 0.20.  
(b) Comparison of spatially averaged c-axis tunneling spectra on Bi-2212
(symbols, from Ref.~\cite{McElroy05,Hudson99}) 
with theoretical calculations (solid lines). Here the Q-values are obtained
by matching the Fermi level of the antibonding band, whereas consideration 
of the bonding band results in similar doping dependent values of 
($V_{\rm CO}$, $\Delta _{\rm SC}$) but different Q-values of 0.384 $\pi$, 
0.379 $\pi$, 0.366 $\pi$ and 0.340 $\pi$ for $\delta$ = 0.10, 0.11, 0.14 and 0.18, 
respectively~\cite{Note}. The fitted $\eta$ value is $\sim 2 \times 10^{-6}$
for $\delta$ = 0.10, 0.11, 0.14 and $\eta \sim 0$ for $\delta$ = 0.18. 
(c) Comparison of a typical tunneling spectrum (open squares)~\cite{ChenCT02} of 
electron-type optimally doped La-112 with theoretical calculations (solid line), 
showing $\Delta _{\rm SC} \sim V_{\rm CO}$ and $\eta \sim 5 \times 10^{-7}$, in contrast 
to $\Delta _{\rm SC} < V_{\rm CO}$ in optimal- and under-doped hole-type cuprates.
(d) Comparison of the break-junction spectra of one-layer electron-type cuprates
PCCO~\cite{Kleefisch01} and LCCO~\cite{Alff03} (symbols) with theoretical calculations 
(solid line), showing $\Delta _{\rm SC} \sim V_{\rm CO}$ for optimally doped PCCO 
and $\Delta _{\rm SC} > V_{\rm CO}$ for underdoped LCCO. The fitted $\eta$ value is 
$10^{-6}$ for both PCCO and LCCO.}
\label{Fig1}
\end{figure}

Using the aforementioned approach, we find that many important features in the quasiparticle 
DOS of both hole- and electron-type cuprates of varying doping levels can be well
described by a set of parameters $( \Delta _{\rm SC}, V _{\rm CO}, \eta , \textbf{Q})$, 
where $\eta$ denotes the magnitude of the quantum phase fluctuations~\cite{ChenCT07,Kwon99a,Kwon01}
and is proportional to the mean-value of $C_{\ell} (\textbf{q})$ in Eq.~(\ref{eq:self})
by the relation $\langle C_{\ell} (\textbf{q}) \rangle = (2 \hbar q/ m_e)^2 \eta$, with $m_e$ being
the electron mass. More specifically, we find that in the event of $V _{\rm CO} > \Delta _{\rm SC}$
and $T = 0$, there are two sets of spectral peak features at $\omega = \pm \Delta _{\rm SC}$ and 
$\omega = \pm \Delta _{\rm eff}$, with the effective excitation gap defined as 
$\Delta _{\rm eff} \equiv (\Delta _{\rm SC} ^2 + V_{\rm CO} ^2)^{1/2}$~\cite{ChenCT07}. The relative
spectral weight of the peak features is largely determined by $\textbf{Q}$, whereas the
magnitude of the zero-bias conductance is sensitive to $\eta$~\cite{ChenCT07}. Moreover, 
the zero-temperature features at $\omega = \pm \Delta _{\rm SC}$ begin to diminish in the spectral
weight and shift to a smaller absolute energy with increasing temperature, and completely vanish above 
$T_c$. In contrast, the features at $\omega = \pm \Delta _{\rm eff}$ evolve with temperature 
into rounded ``humps'' at $\omega \sim \pm V _{\rm CO}$ for $T >\sim T_c$~\cite{ChenCT07}, 
resulting in the well known low-energy pseudogap (PG) phenomena in optimally doped and underdoped 
hole-type cuprates. On the other hand, if $V _{\rm CO} < \Delta _{\rm SC}$, only one set of 
zero-temperature spectral peak features can be resolved at $\omega = \pm \Delta _{\rm eff}$, 
so that no PG is observed above $T_c$, which is consistent with the general findings 
in electron-type cuprates~\cite{ChenCT07}.  

As exemplified in Fig.~\ref{Fig1}, we compare the quasiparticle tunneling spectra 
taken on four different families of cuprates with the calculated DOS, where 
we have used different combinations of coexisting SC and CO phases based on established 
empirical facts: $d_{x^2-y^2}$- or $(d_{x^2-y^2}+s)$-wave SC and disorder-pinned 
SDW for hole-type cuprates $\rm YBa_2Cu_3O_x$ (Y-123)~\cite{Yeh01,Wei98} and 
$\rm Bi_2Sr_2CaCu_2O_x$ (Bi-2212)~\cite{Tsuei00}, $s$-wave SC and CDW for the infinite-layer
electron-type cuprate $\rm Sr_{0.9}La_{0.1}CuO_2$ (La-112)~\cite{ChenCT02}, and 
$d_{x^2-y^2}$-wave SC and disorder-pinned SDW in one-layer electron-type cuprate
superconductors $\rm Pr_{1.85}Ce_{0.15}CuO_{4-y}$ (PCCO) and $\rm La_{1.884}Ce_{0.116}CuO_{4-y}$ 
(LCCO)~\cite{Tsuei00}. In all cases, we have employed realistic bandstructures 
and Fermi energies for given cuprates under consideration. 

\begin{figure}
\begin{center}
\includegraphics*[keepaspectratio=1,height=3in]{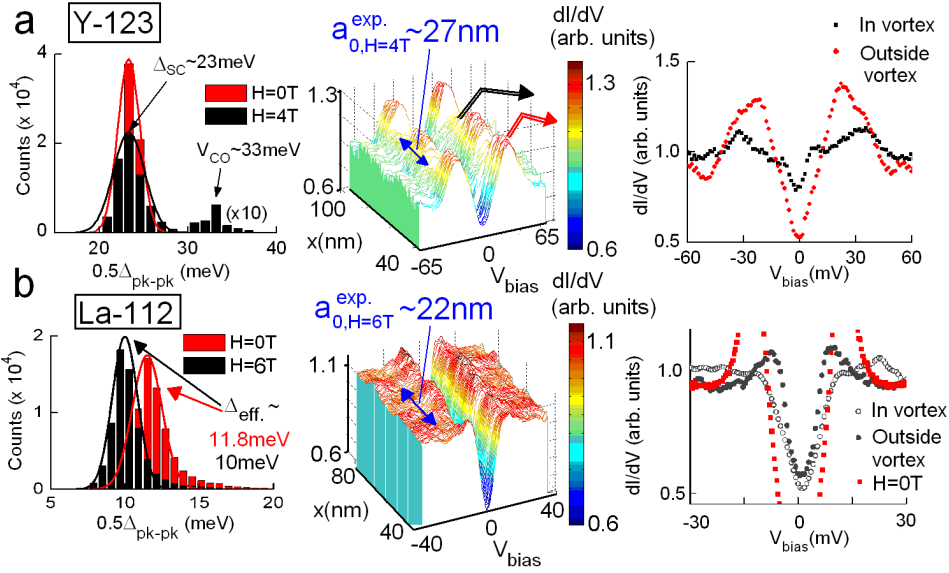}
\end{center}
\caption{(Color online) Spatially resolved quasiparticle spectra of cuprate superconductors: 
(a) Left panel, zero-field histogram of $\Delta_{\rm SC}$ (red or light grey) and
finite-field ($H = 4$ Tesla) histogram of $\Delta_{\rm SC}$ and $V_{\rm CO}$ (dark grey) taken 
from an optimally doped Y-123 over an $\rm (100nm \times 100nm)$ area at $T = 9$ K. Here
the empirical gaps were taken from one-half of the peak-to-peak energy values of the tunneling spectra, 
so that in zero field only $\Delta_{\rm SC}$ were recorded whereas the ``hump'' features associated with 
$\Delta_{\rm eff}$ were not registered. In contrast, at $H = 4$ Tesla the empirical gap values identified from 
one-half of the peak-to-peak values revealed two sets of values. One of the gap values associated 
with the coherence peaks outside of the vortex core was found to be consistent with the zero-field 
$\Delta_{\rm SC}$, whereas the other associated with the PG-like features inside the vortex core was 
consistent with the $V_{\rm CO}$ value derived from zero-field spectral fitting.
Hence, we find a homogeneous $\Delta_{\rm SC} = 23 \pm 1$ meV at $H$ = 0 and a broader distribution 
of $\Delta_{\rm SC} = 23 \pm 2$ meV and $V_{\rm CO} = 33 \pm 2$ meV 
at $H = 4$ Tesla. The revelation of the CO histogram is consistent with the suppression of SC 
inside each vortex core and the fact that $V_{\rm CO} > \Delta _{\rm SC}$.  
Central panel, spatial evolution of the vortex-state quasiparticle spectra of Y-123 taken under
a c-axis magnetic field $H = 4$ Tesla. The periodic spatial variations of the quasiparticle 
spectra are consistent with a vortex lattice constant $a_0 \sim 27$ nm. Right panel, comparison of
the spectrum at the center of a vortex core with that outside of the vortex core, the former
exhibit PG-like features at $V_{\rm CO} = 33 \pm 2$ meV 
upon the suppression of the SC coherence peaks at $\Delta_{\rm SC} = 23 \pm 2$ meV.
(b) Left panel, zero-field histogram of $\Delta_{\rm eff}$ (Red or light grey) and finite-field 
($H = 6$ Tesla) histogram of $\Delta_{\rm eff}$ (dark grey) taken from an optimally doped La-112 
over an $\rm (100nm \times 100nm)$ area at $T = 9$ K, showing a homogeneous $\Delta_{\rm eff} = 11.8 \pm 1.5$ 
meV at $H$ = 0 and a downshift to $\Delta_{\rm eff} = 10 \pm 1$ meV at $H = 6$ Tesla 
due to the overall suppresion of $\Delta _{\rm SC}$ by magnetic field. Here we note that 
the c-axis upper critical field of La-112 is $H_{c2}^{c} \sim 12$ Tesla~\cite{Zapf05}.
Central panel, spatial evolution of the vortex-state quasiparticle spectra of La-112 taken under
a c-axis magnetic field $H = 6$ Tesla. The periodic spatial variations of the quasiparticle 
spectra are consistent with a vortex lattice constant $a_0 \sim 22$ nm. Right panel, quasiparticle
spectrum at the center of a vortex core (black open squares) and that outside of the vortex core 
(black solid squares) for $H = 6$ Tesla are compared with that for $H = 0$, showing both decreasing 
$\Delta _{\rm eff}$ with increasing field and the revelation of PG-like features 
in the center of a vortex core.}
\label{Fig2}
\end{figure}

The data in Fig.~\ref{Fig1}(a), represented by symbols, are our
c-axis tunneling spectra on Y-123 with varying doping levels~\cite{Yeh01,Wei98}. 
The c-axis tunneling data in Fig.~\ref{Fig1}(b) on $\rm Bi_2Sr_2CaCu_2O_x$ (Bi-2212), 
shown as symbols for four different nominal hole-doping levels, are taken from 
Refs.~\cite{McElroy05,Hudson99}. In Y-123 and La-112,
the quasiparticle spectra exhibit long-range spatial homogeneity so that the bulk doping level 
is representative of the local doping level, as manifested by the zero-field histograms 
of $\Delta _{\rm SC}$ and $\Delta _{\rm eff}$ in Figs.~\ref{Fig2}(a)-(b). 
In contrast to Y-123 and La-112, caution must 
be taken in studying the quasiparticle spectra of Bi-2212 because of the strong spatial 
inhomogeneity in the latter~\cite{McElroy05}. We estimate the local doping level of 
Bi-2212 by correlating the dominant spatially averaged spectrum of each sample with 
its bulk doping level, so that each empirical spectrum shown in Fig.~1(b) is spatially 
averaged. The data in Fig.~\ref{Fig1}(c) are taken from a representative 
spectrum amongst a set of momentum-independent quasiparticle tunneling spectra
of the optimally doped La-112 with $T_c \approx 43$ K~\cite{ChenCT02}, and the 
data in Fig.~\ref{Fig1}(d) for optimally doped PCCO and overdoped LCCO are taken from
break-junction spectra in Refs.~\cite{Kleefisch01,Alff03}.

To further verify the scenario of coexisting COs and SC in cuprate superconductors, we 
perform spatially resolved vortex-state quasiparticle spectra on two cuprates that exhibit
homogeneous zero-field tunneling spectra, the optimally doped Y-123 and La-112. In 
conventional superconductors with SC being the only ground state, large supercurrents 
inside the vortex core are known to suppress SC, leading to either bound states 
with enhanced conductance at zero bias or flat conductance in the quasiparticle 
spectra~\cite{Fischer07}. In contrast, various studies of the vortex-state 
quasiparticle tunneling spectra of hole-type cuprate superconductors have revealed
PG-like features inside the vortex core~\cite{Fischer07}. Our own investigation on
Y-123 and La-112 also shows similar behavior in both hole- and electron-type cuprates,
as illustrated in Figs.~\ref{Fig2}(a)-(b). Furthermore, we note that the energy scale
of the PG-like features at the center of the vortex core is consistent with the
$V_{\rm CO}$ value derived from our theoretical fitting to the zero-field experimental 
data, as shown in the histograms and the vortex-state conductance maps of both Y-123 
and La-112 in Figs.~\ref{Fig2}(a)-(b) and also in Figs.~\ref{Fig1}(a) and (c). 
Hence, we suggest that the anomalous vortex-core quasiparticle spectra in cuprate 
superconductors can be understood in terms of a remaining CO upon the suppression of SC.

\begin{figure}
\begin{center}
\includegraphics*[keepaspectratio=1,height=2.4in]{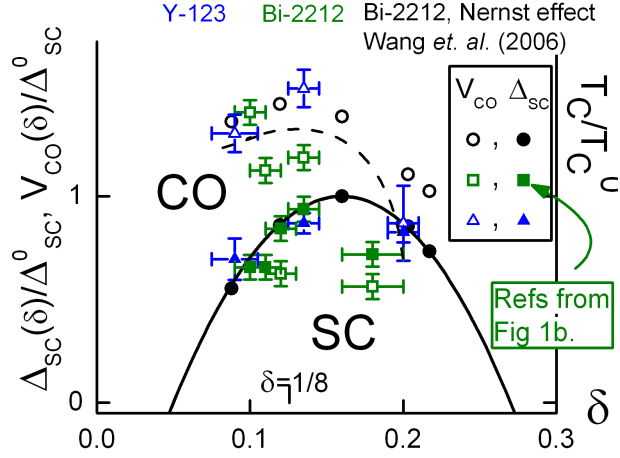}
\end{center}
\caption{(Color online) Doping dependence of $\Delta _{\rm SC}$, $V_{\rm CO}$ and 
$T_c$ in hole-type cuprates Y-123 (blue or dark triangles) and Bi-2212 (green or light 
squares), where $\Delta _{\rm SC}$ and $V_{\rm CO}$ are normalized 
to their corresponding SC gaps at the optimal doping, $\Delta _{\rm SC} ^0$, 
and $T_c$ is normalized to the value at the optimal doping $T_c^0$. 
The doping dependent $T_{\rm onset} (\delta)$ for the onset of diamagnetism 
and the Nernst effect in Bi-2212 together with the corresponding $T_c (\delta )$ 
from Ref.~\cite{WangY05} are shown in black circles for comparison.}
\label{Fig3}
\end{figure}

Having established the dependence of spectral characteristics on $\Delta _{\rm SC}$ 
and $V_{\rm CO}$, we consider next the effect of the CO wave-vector $\textbf{Q}$ as
a function of the doping level. We note that in the data fitting (solid lines) 
shown in Fig.~\ref{Fig1}(a)--(d), we have assumed that $\textbf{Q}$ satisfies the 
nesting conditions $|\textbf{k}+\textbf{Q}| \sim k_F$ and $|\textbf{k}| \sim k_F$ so that 
the quasiparticle excitations only occur near the Fermi momentum $k_F$. This assumption 
is justifiable for the hole-type cuprates because the degree of incommensurate 
spin fluctuations in these cuprates correlates with the doping level~\cite{Wells97}. 
Thus, we derive the doping-dependent parameters $\Delta _{\rm SC} (\delta)$ and 
$V_{\rm CO} (\delta)$ for different cuprates by fitting curves to experimental data, 
and the parameters $\Delta _{\rm SC} (\delta)$ and $V_{\rm CO} (\delta)$ normalized 
to the SC gap at the optimal doping level ($\Delta _{\rm SC} ^0$) of each cuprate 
family are summarized in Fig.~\ref{Fig3}, together with the normalized SC transition 
temperature ($T_c/T_c ^0$) and the onset temperature for diamagnetism and the Nernst 
effect~\cite{WangY05} ($T_{\rm onset}/T_c ^0$). Our theoretical fitting to the quasiparticle 
DOS not only captures the primary low-energy features of the tunneling spectra 
in Fig.~\ref{Fig1}(a)--(d) but also yields a doping dependent $\Delta _{\rm SC} (\delta)$ 
that closely resembles the doping dependence of $T_c (\delta)$. In contrast, $V_{\rm CO} (\delta)$ 
generally increases with decreasing doping level in the doping range $0.1 < \delta < 0.22$, 
and the overall doping dependence also follows that of $T_{\rm onset} (\delta)$ 
for diamagnetism and the Nernst effect in Bi-2212~\cite{WangY05}, as shown in Fig.~\ref{Fig3}. 
We further remark that the values of $\Delta _{\rm SC}$ and $V_{\rm CO}$ derived 
from fitting the quasiparticle DOS are insensitive to small variations in $\textbf{Q}$, 
whereas ARPES characteristics are more dependent on $\textbf{Q}$. 

In addition to accounting for the primary low-energy features of doping dependent 
quasiparticle tunneling spectra, the notion of coexisting COs and SC can explain the 
spatially varying local density of states (LDOS) and the corresponding Fourier 
transformation of the LDOS (FT-LDOS) in Bi-2212~\cite{ChenCT03,McElroy05,Hoffman02}. 
That is, the spatially varying LDOS may be attributed to varying parameters 
$(\Delta _{\rm SC}, V_{\rm CO})$ so that the LDOS ${\cal N} (\textbf{r}, \omega)$ 
is position dependent. In particular, $V_{\rm CO} (\delta)$ exhibits much stronger doping 
dependence than $\Delta _{\rm SC} (\delta)$, as shown in Fig.~\ref{Fig3}. Therefore, 
the primary cause of spatially inhomogeneous LDOS in Bi-2212 (in contrast to those in
Y-123 and La-112) may be attributed to varying $V_{\rm CO}$ due to varying doping levels, 
leading to broader peaks at quasiparticle energies $\omega \sim \pm \Delta _{\rm eff}$~\cite{ChenCT07}. 
The spatially varying $V_{\rm CO}$ can give rise to quasiparticle scattering, 
yielding FT-LDOS that contains information about quasiparticle interference and 
the presence of CO~\cite{ChenCT03}.

\begin{figure}
\begin{center}
\includegraphics*[keepaspectratio=1,width=5in]{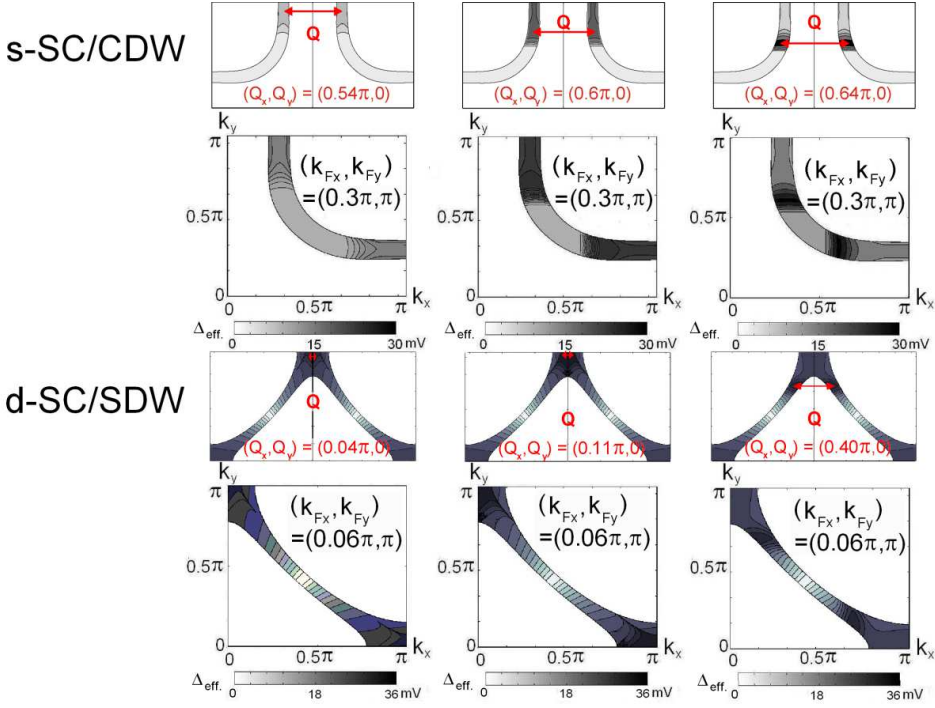}
\end{center}
\caption{(Color online) Competing order-induced dichotomy in the momentum-dependent
effective gap $\Delta _{\rm eff} (\textbf{k})$ is illustrated in 
the first quadrant of the BZ: The second row corresponds to the $\Delta _{\rm eff} (\textbf{k})$
map for $s$-wave SC coexisting with CDW ($s$-SC/CDW), and the fourth row corresponds to 
the $\Delta _{\rm eff} (\textbf{k})$ map for $d_{x^2-y^2}$-wave SC coexisting with SDW ($d$-SC/SDW). 
The wave-vector $\textbf{Q}$ of the CO along either $(\pi ,0)$ or $(0, \pi )$ 
direction varies from $|\textbf{Q}| < 2k_F$ in the left panels to 
$|\textbf{Q}| \sim 2k_F$ in the middle panels and to $|\textbf{Q}| > 2k_F$ 
in the right panels. The phase space associated with the CDW (disorder-pinned SDW) 
contributions to the $s$-wave ($d_{x^2-y^2}$-wave) SC for different $|\textbf{Q}|$-values 
is shown in the first (third) row, and the bandstructures for these calculations 
have included the bilayer splitting effects~\cite{Schabel98,Andersen95,Hoogenboom03}. 
The $|\textbf{Q}|$-values shown in the third row for the anti-bonding band
correspond to twice of those of the SDW.}
\label{Fig4}
\end{figure}

Next, we consider the situation when the nesting condition for $\textbf{Q}$ is relaxed. 
As exemplified in Fig.~\ref{Fig4}, we compare the effective gap ($\Delta _{\rm eff}$) 
in the first quadrant of the Brillouin zone (BZ) for $s$-wave SC with coexisting CDW 
(second row) and for $d_{x^2-y^2}$-wave SC with coexisting disorder-pinned SDW (fourth row) 
assuming different $\textbf{Q}$-values, where the coupled quasiparticle states 
in the first BZ are illustrated above each $\Delta _{\rm eff}$ plot. For $\textbf{Q}$ varying 
from $|\textbf{Q}| < 2k_F$ (left panels), $|\textbf{Q}| \sim 2k_F$ (middle panels), 
to $|\textbf{Q}| > 2k_F$ (right panels), we find strongest CO-induced effects
for $|\textbf{Q}| \sim 2k_F$, implying maximum impact of the 
CO on the ground state and the low-energy excitations of the cuprates if the CO 
wave-vector is correlated with $k_F$. Interestingly, if we consider $s$-wave SC coexisting with 
a commensurate CDW ($|\textbf{Q}| \sim 2 \pi /3 > 2k_F$), 
the resulting $\Delta _{\rm eff} (\textbf{k})$ becomes maximum near the ``hot spots'' 
({\it i.e.}, the $\textbf{k}$-values where the antiferromagnetic BZ and the Fermi 
surface intercept) of the optimally doped electron-type cuprates, as shown
in the right panel of the second row in Fig.~\ref{Fig4}. This finding is analogous to the 
ARPES data obtained on electron-type $\rm Pr_{0.89}LaCe_{0.11}CuO_4$~\cite{Matsui05},
where a momentum-dependent excitation potential with maximum magnitude near the hot spots 
is inferred and attributed to quasiparticle coupling with background antiferromagnetism (AFM). 
However, the existence of a long-range AFM order in zero fields or a field-induced 
magnetic order in the SC state of electron-type cuprates remains inconclusive~\cite{Motoyama06,Dai05}. 
Hence, our conjecture of $s$-wave SC with a commensurate CDW provides an alternative explanation for the 
observation in Ref.~\cite{Matsui05}. 

\begin{figure}
\begin{center}
\includegraphics*[keepaspectratio=1,height=4in]{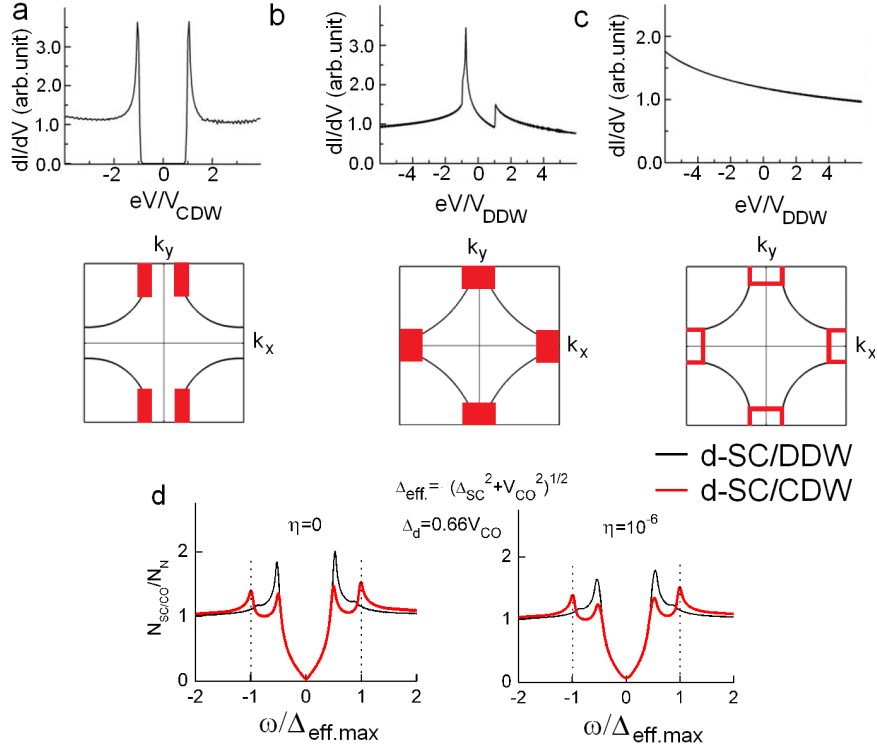}
\end{center}
\caption{(Color online) Comparison of CDW and DDW contributions 
to the quasiparticle excitation spectra of hole-type cuprates in the
one-band approximation: (a) Quasiparticle DOS due to pure CDW in an optimally 
doped hole-type cuprate. The phase space associated with the CDW contributions 
in the first BZ is indicated (red or dark gray bars) in the lower panel. (b) Quasiparticle 
DOS due to pure DDW under a nearly nested condition~\cite{Bena04}. The phase space
associated with DDW in the first BZ is shown in the lower panel. (c) Quasiparticle 
DOS due to pure DDW in an optimally doped hole-type cuprate with realistic 
bandstructures~\cite{Hoogenboom03}. The Fermi surface shown in the lower panel
reveals a small phase space associated with DDW (red or dark gray lines). 
Here we note that related findings for the absence of gapped features in the DOS of 
pure DDW have also been discussed in Ref.~\cite{Morr02}.
(d) Comparison of the quasiparticle spectra calculated for coexisting 
$d_{x^2-y^2}$-wave superconductivity and DDW ($d$-SC/DDW) and coexisting $d_{x^2-y^2}$-wave 
superconductivity and CDW ($d$-SC/CDW) at $T = 0$, with the same bandstructure parameters used 
in (c). The left panel corresponds to mean-field results with $\eta = 0$, and the 
right panel corresponds to the spectra with finite quantum fluctuations $\eta = 10^{-6}$. 
The $d$-SC/DDW spectra only reveal one set of peaks associated with SC. In contrast, the 
$d$-SC/CDW spectra yield both the SC coherence peaks at $\pm \Delta _{\rm SC}$ and the satellite
features at $\pm \Delta _{\rm eff}$, the latter evolving into the low-energy PG at 
$T > T_c$~\cite{ChenCT07}.}
\label{Fig5}
\end{figure}

To examine the effect of DDW on the cuprates, it is informative to make
comparison with the CDW case. As illustrated in Fig.~\ref{Fig5}(a), we find that in the 
one-band approximation the phase space associated with DDW is more restrictive so 
that the resulting quasiparticle spectra only exhibits PG-like features for a nearly 
half-filling condition (see Fig.~\ref{Fig5}(b))~\cite{Bena04}. If we consider the 
Fermi surface of a realistic hole-type cuprate~\cite{Hoogenboom03} with a doping 
level deviating from half-filling, we find that due to the small phase space 
associated with the DDW, the DDW contributions to the quasiparticle low-energy 
excitations does not yield the gap-like features commonly observed in experiments, 
as shown in Fig.~\ref{Fig5}(c). If we take the bilayer splitting into 
consideration for Bi-2212 and Y-123, we find that the the anti-bonding band 
in Bi-2212 is comparable to the nearly nested condition, whereas neither the 
bonding nor antibonding band of Y-123 matches the nested condition~\cite{Schabel98}. 
The incompatibility of DDW with SC is also consistent 
with recent numerical studies of a two-leg ladder system, which reveal mutually 
exclusive DDW and $d_{x^2-y^2}$-wave SC states with realistic physical
parameters~\cite{Schollwock03}. 

To further contrast the compatibility of CDW (or disorder-pinned SDW) 
and the incompatibility of DDW with cuprate SC, we illustrate in Fig.~\ref{Fig5}(d) 
comparison of the quasiparticle spectra calculated for coexisting $d_{x^2-y^2}$-wave SC 
and DDW ($d$-SC/DDW) and coexisting $d_{x^2-y^2}$-wave SC and CDW ($d$-SC/CDW), 
using the same bandstructure parameters employed in Fig.~\ref{Fig5}(c). The left panel 
of Fig.~\ref{Fig5}(d) corresponds to mean-field results ($\eta = 0$), and the 
right panel corresponds to the spectra with finite quantum fluctuations 
$\eta = 10^{-6}$. Clearly for both cases the $d$-SC/DDW spectra 
only reveal one set of peaks associated with SC, whereas the $d$-SC/CDW spectra 
can account for both the SC coherence peaks at $\pm \Delta _{\rm SC}$ and the 
PG features at $\pm \Delta _{\rm eff}$, suggesting that CDW (or disorder-pinned SDW) 
is a more likely CO than DDW for the low-energy PG phenomenon.

In the context of relevant CO's for cuprate superconductors, it is worth noting a recent 
development that reports the onset of finite Kerr signals, albeit very small, below the PG temperature 
in Y-123 and in the absence of magnetic fields~\cite{Xia07}. This finding indicates the occurrence 
of ferromagnetic-like signals upon the PG formation, and is compatible with broken time-reversal 
symmetry. Interestingly, various CO's such as AFM, SDW, DDW and the triangular current-loop 
model~\cite{Varma97} are all consistent with broken time-reversal symmetry, but are at the same 
time incompatible with the existence of ferromagnetism. On the other hand, local spontaneous 
magnetic moments solely associated with local impurity phases in Y-123 
have been reported from scanning SQUID microscopy studies~\cite{Tafuri00}, which may give rise to
weak ferromagnetic signals on average. Although the physical origin of the onset of Kerr
signals below the PG temperature remains inconclusive, the aforementioned empirical and numerical 
information at least suggests that the DDW order parameter is unlikely the primary CO responsible 
for the low-energy PG in cuprate superconductors. 

\section{Discussion}
\label{Discussion}

In hole-type cuprates, additional high-energy satellite features in the 
quasiparticle spectra are known to exist, and the corresponding characteristic 
energies in Bi-2212 have been attributed to magnetic excitations~\cite{Chubukov00}. 
Recently, evidence for phonon modes at an energy $\omega > \Delta _{\rm SC}$ and
in-between the ``dip'' and ``hump'' spectral features of Bi-2212 has also been 
identified~\cite{LeeJ06}. However, whether these high-energy bosonic modes are 
relevant to the occurrence of the high-energy PG remains an open issue 
for investigation.

If the high-energy PG is indeed associated with a bosonic energy scale
$V_{\rm PG}$, we may construct a generic temperature ($T$) vs. 
doping level ($\delta$) phase diagram of the cuprates in terms of the interplay 
of three primary energy scales: $V_{\rm PG}$, $V_{\rm CO}$ and $\Delta _{\rm SC}$,
which correspond to temperature scales of $T_{\rm PG} (\delta)$, $T^{\ast} (\delta)$ 
and $T _c (\delta)$. In the case of hole-type cuprates, generally 
$V_{\rm CO} > \Delta _{\rm SC}$ for a wide range of doping levels, so that the CO occurs 
at $T^{\ast} (\delta)> T_c (\delta)$. We speculate that the larger $V_{\rm CO}$ 
in hole-type cuprates may be attributed to an enhanced charge transfer along the Cu-O 
bonding through significant coupling of the conduction carriers to the longitudinal optical 
(LO) phonons~\cite{Tachiki03}. In contrast, no enhanced charge transfer can occur
through the LO phonons in the electron-type cuprates so that $V_{\rm CO}$ is generally 
smaller. As a result, there is no apparent low-energy PG associated with the electron-type 
cuprates if $H = 0$~\cite{Kleefisch01,ChenCT02}. On the other hand, the higher-energy 
PG exists in both electron- and hole-type cuprates, which may be related to spin 
fluctuations and thus $V_{\rm PG} \gg \Delta _{\rm SC}$. 
Finally, we note that in the language of the slave-boson theory~\cite{LeePA06}, 
$V_{\rm PG}$ may be thought of as the spinon PG in the underdoped limit. 
In this context, we may consider the spinon PG phase determined by the 
energy scale $V _{\rm PG}$ as the highly degenerate ``parent phase'' 
of all cuprates, so that AFM, SC, and CO are symmetry-breaking instabilities 
derived from this spin-liquid like parent phase~\cite{Yeh07}. 

\section{Conclusion}
\label{Conclusion}

In summary, we have investigated the low-energy excitations of cuprate superconductors 
in the context of coexisting COs and SC, with special emphasis on examining the
effect of varying doping levels ($\delta$) and CO wave-vectors ($\textbf{Q}$).
For various hole- and electron-type cuprate superconductors, the doping dependence 
of the CO energy $V_{\rm CO} (\delta)$ derived from fitting zero-field experimental 
tunneling spectra is consistent with the doping dependence of the low-energy PG and 
of the onset temperature for diamagnetism and the Nernst effect~\cite{WangY05}, which
increases with decreasing $\delta$. Moreover, $V_{\rm CO}$ values derived from
the zero-field tunneling spectra of optimally doped Y-123 and La-112 are found to
be consistent with the PG-like energy scales found in the center of the vortex core
where SC is nearly fully suppressed. On the other hand, the SC gap $\Delta_{\rm SC}(\delta)$ 
derived from zero-field quasiparticle spectra is found to scale with $T_c (\delta)$, 
and the condition $V_{\rm CO} (\delta) > \Delta _{\rm SC} (\delta)$ 
holds for under- and optimal doping levels. In addition, the wave-vector $\textbf{Q}$ 
of the CO in hole-type cuprates appears to be incommensurate and doping dependent, 
whereas the condition $V_{\rm CO} (\delta) \le \Delta _{\rm SC} (\delta)$ is found in 
electron-type cuprate superconductors and the corresponding $\textbf{Q}$ 
appears to be commensurate. Finally, for realistic bandstructures DDW does not 
couple well to the low-energy quasiparticle excitations of doped hole-type 
cuprates, and is therefore not a favorable CO responsible for the low-energy 
PG phenomena in zero fields.


\begin{thebibliography}{44}
\bibitem{Note}
The doping levels listed here have been adjusted slightly from those given in
Ref.~\cite{McElroy05} by better matching to the bulk $T_c$ values.
\end{thebibliography}
\end{document}